%% file: Main.tex
\documentclass[conference]{IEEEtran}
\IEEEoverridecommandlockouts
\usepackage{comment}
\usepackage{verbatim} 
\usepackage{rotating}
\usepackage{hyperref}

\input{Settings/packages.tex}

\begin{document}

\title{When Interactive Graphic Storytelling Fails}

\author{
    \IEEEauthorblockN{James Barela\IEEEauthorrefmark{1},
    Tiago Gasiba\IEEEauthorrefmark{2}, 
    Santiago Reinhard Suppan\IEEEauthorrefmark{3},
    Marc Berges\IEEEauthorrefmark{4},
    Kristian Beckers\IEEEauthorrefmark{5}
    }
    \IEEEauthorblockA{
      \IEEEauthorrefmark{1}  \IEEEauthorrefmark{4}  Technische Universit{\"a}t M{\"u}nchen~~~~
      \IEEEauthorrefmark{2}   \IEEEauthorrefmark{3}   \IEEEauthorrefmark{5}  Siemens AG, M{\"u}nchen\\
      \IEEEauthorrefmark{1}
      james.c.barela@tum.de
        \IEEEauthorrefmark{2}
      tiago.gasiba@siemens.com, \\
        \IEEEauthorrefmark{3}
      Santiago.Suppan@Siemens.com,
       \IEEEauthorrefmark{4}
      Berges@TUM.de,
       \IEEEauthorrefmark{5}
      kristian.beckers@siemens.com
          }
    }

\maketitle

\begin{abstract}
Many people are unaware of the digital dangers that lie around each cyber-corner. Teaching people how to recognize dangerous situations is crucial, especially for those who work on or with computers. 
We postulated that interactive graphic vignettes could be a great way to expose professionals to dangerous situations and demonstrate the effects of their choices in these situations. In that way, we aimed to inoculate employees against cybersecurity threats.

We used the Comic-BEE platform to create interactive security awareness vignettes and evaluated for how employees of a major industrial company perceived them. For analysing the potential of these comics, we ran an 
evaluation study as part of a capture-the-flag (CTF) event, a interactive exercise for hacking vulnerable software. We evaluated whether the comics fulfilled our requirements based on the responses of the participants.
We showed the comics, on various cybersecurity concepts, to 20 volunteers. In  the context of a CTF event, our requirements were not fulfilled. Most participants considered the images distracting, stating a preference for text-only material. 


\end{abstract}


\input{1-intro.tex}



\input{x-RW_BG.tex}

\input{3-approach.tex}

\input{4-results.tex}

\input{5-conclusion.tex}

\section*{Acknowledgements}
 The authors would like to thank all participants of the experiment for their time and their valuable insights and suggestions. The authors would also like to thank the providers of the comic platform. 
\bibliographystyle{IEEEtran}
\footnotesize
\bibliography{CTF_Attempt_v02,re2016,bibliography}

\end{document}

%% file: Settings/packages.tex
\usepackage{booktabs} 
\usepackage{url}
\usepackage{listings}
\usepackage{caption,setspace}
\usepackage{siunitx,threeparttable}
\usepackage[english]{babel}
\usepackage{float}
\usepackage{floatflt}
\usepackage{fancyhdr}
\usepackage{color}
\usepackage{mdwlist}
\usepackage{paralist}
\usepackage{array}
\usepackage{longtable}
\usepackage{listings}
\usepackage[utf8]{inputenc}
\usepackage{rotating}
\usepackage{ragged2e}
\usepackage{tcolorbox}
\usepackage[autostyle]{csquotes} 
\usepackage{soul}

\usepackage{setspace} 
\usepackage{etoolbox}
\AtBeginEnvironment{quote}{\singlespacing\small}

\usepackage[colorinlistoftodos]{todonotes}
\usepackage{enumitem}

%% file: 1-intro.tex
\section{Introduction}

Our world has become vastly dependent on information technology. Millions of devices communicate every second in cyberspace. 
We are now connected to one another to a degree that seemed inconceivable just 20 years ago, but with so many connections, we are confronted with just as much, if not more, vulnerability; 
Hubbard \& Seiersen,
in their book \emph{How to Measure Anything in Cybersecurity Risk} \cite{Hubbard2016},
lay out the case that the global attack surface is increasing from at least four perspectives: (1) the  number of people on the Internet, (2) the number of online resources that each person is consuming, (3) the vulnerabilities that come with those people and resources, and (4) the risk from building online services on top of one another, which could result in a ``breach cascade". 
Modern industrial systems and critical infrastructure should account for cyber threats, especially when corporate organizations, national economies, and public safety have been put at risk \cite{Newhouse2017,Blythe2013}. 
As a result of these omnipresent risks, laws and standards have been created, such as NERC CIP-003-7, 
IEC 62443, 
and ISO 27001,
that attempt to mitigate the risks posed by cyberattacks.
These international standards acknowledge that among the weakest points in our security are not the technologies themselves, but the \emph{people} behind the technologies. 
People often lack the knowledge and training to avoid even the simplest of hacker schemes, such as social engineering, or phishing attacks.
Despite these requirements and all the best efforts of IT Security departments, organizations are still only as strong as the employees who regularly engage with the technologies.
People are the front line of defense against cyber threats---in particular, the engineers, managers, and other staff who interact directly with information systems and their security measures on a daily basis. 
Therefore, these international standards require that such personnel are well-trained, that they are aware of the risks, and that they are prepared and have the resources to mitigate those risks.

Unfortunately, this is as far as mandates, requirements, and laws can go; 
companies have to interpret and implement them, as well as measure if the training yields the desired results.
If there is an attack, and it is not averted because employees did not receive the right training, then it will have already been too late.

The question is, how do we train companies' most valuable resource, their employees? 
And how do we ensure that the training has been efficient and cost-effective? 
National institutions exist that are dedicated to the sole purpose of education (the Department of Education, USA; Bundes Ministerium f{\"u}r Bildung und Forschung, Germany; etc).
But what this work needs to take into consideration is that the people in need of educating are (1) in the Age of Information, a different world from anything that has ever existed before; and (2) that they are not children being prepared for the workplace, but adults who are already \emph{in} the workplace. The old paradigms are falling out of favor and it would be prudent to take advantage of the resources that are available, i.e., computers and the Internet. 
Security training needs to be close to the context employees work in. 
Therefore, our training content was based on interviews with security experts working in the company itself. Training also needs to be cost-effective and scale well, which is why we selected an online platform. 


In addition, understanding computers to be the main method for delivery (as opposed to lectures or in-person presentations) the media should take full advantage of the properties of a computer. Given Mayer's extensive work on learning with multimedia, this paper will explore a new method, which shows promise for training employees effectively by reducing cognitive load \cite{Mayer2003,Mayer2002}.
Our method uses a vignette modality so that participants had to further engage with the comic in a way that implements situated learning \cite{Brown1988}. Furthermore, we evaluate the claims made by the developers of the comic-creation platform, especially with regard to the time and skills that are required to create a comic.\footnote{The following will be used interchangeably: comic, story, interactive stories, graphical vignettes, etc}

Our requirements for providing an online educational tool for cybersecurity topics are: 
\begin{itemize}[leftmargin=+.48in]
    \item[ {\bf REQ1:}] Trainees shall be entertained during the course of the training
    \item[ {\bf REQ2:}] The comics shall not be perceived as disturbing or even cause stress during game play
    \item[ {\bf REQ3:}] Trainees shall recognize the context and feel compelled to show they know the right answer 
    \item[ {\bf REQ4:}] Trainees shall understand how to use the comics intuitively 
\end{itemize}

In short, we present and evaluate a method for developing educational cybersecurity comics and then using them to train employees in a company.
We present the method of development and the subsequent evaluation, which was conducted as part of a capture the flag event, with 20 employees of a major industry player. In general, the comic approach did not fulfill our requirements as part of a CTF, we think that the lessons learned can and should be shared with the scientific community.
We close the paper with the participants' reasoning, critical discussion on the results, and practical advice. 


%% file: x-RW_BG.tex
\section{Related Work and Background}

Security training has often been perceived as an uninteresting and even boring topic. Therefore, several researchers have been developing serious games to combine entertainment and training in this field. 
\textit{CyberCIEGE}~\cite{irvine2005cyberciege} is a role playing video game, where players act as an information-security decision-maker of a company. Players have to minimize risk, while continuing to work. Using the typical mechanics of a board game, \textit{PlayingSafe} ~\cite{newbould2009playing} consists of multiple choice questions about cybersecurity.
Another game, \textit{SEAG}~\cite{olanrewajusocial} also uses multiple choice questions, and additionally players have to match cybersecurity terms with their respective pictures. Our work differs from these, since none of them use interactive comic-based vignettes that are based on real-world, industry concerns.

Comics have been used before to teach about complex issues, such as training soldiers on matters of military leadership~\cite{Gordon2006}. 
These comics displayed a problem and participants would fill in a response. Responses could then be discussed on a forum that included responses from previous participants. Our work does not utilize forums, and the interaction is of the multiple-choice type. This allows for more immediate participant feedback.

In regards to creating or generating comics, Microsoft has done work on automatically generated comics based on cursory language analysis~\cite{Kurlander1996}. Their goal was not to produce training material, but to visually represent chat environments to enrich the user experience. 

Ledbetter~\cite{Ledbetter2016} stated that the goals for creating Comic-BEE are: to introduce young people to cybersecurity content in fun and engaging ways, and thereby encourage them to contemplate and pursue cybersecurity career paths. Among the reasons they cite are that (1) educational comics can be appealing on multiple levels, including engaging and increasing interest in readers; and (2) young people are not being sufficiently exposed to  cybersecurity concepts, despite being ever more surrounded by technology. 

This paper is concerned with how to design instructional material, and how it can be applied to a company and other institutions like it.
The evolution of instructional design has been written about by Wilson~\&~Cole~\cite{Wilson1996}; especially as it can be understood through the lens of cognitive models: from the 1960's, with Behavioral psychology; through Information Processing psychology in the 1970's and 1980's; to the 1990's and the present, which emphasizes the construction of knowledge and the role of social mediation.
The state-of-the-art is now considered to be ``Constructivism," which has been the predominant learning theory over the past three decades~\cite{Fosnot1996}.
Constructivism presupposes two main tenets: (1) knowledge is a construction that comes from an active interaction with the world, and (2) that it is inherently an adaptive process that is not necessarily concerned with the ontological nature of reality. This will be important, as our method utilizes simulacra of real-world cybersecurity concepts.

%% file: 3-approach.tex
\section{Creating our Interactive Graphical Stories}

To begin this study, the authors sought a platform that was designed for creating educational comics in a quick and easy way.
Learning material was then generated that was relevant to the working environment of the target participants.
The information to create the training material was gathered from interviews with security experts from the industry.
The gathered information was consolidated and concentrated into 5 main topic areas, each of which would become its own comic.
The developed comics were deployed along with a survey to gather qualitative data about how the comics were received by the participants.


\subsection{The Comic Development Platform}
\label{sec:comicplatform}

The comic platform is a tool developed by Secure Decisions, a division of Applied Visions, Inc., which has been sponsored by the Science and Technology division of Homeland Security in the United States.\footnote{Contract \#HHSP233201600057C, retrieved from https://govtribe.com/vendor/applied-visions-inc-northport-ny}



This platform was chosen as it is aligned with the goals of this work and also allows the implementation of the requirements defined in this paper. The Comic-BEE team was contacted and they provided the requisite credentials for creating comics on their site.
The platform was evaluated with a first attempt to create a small sample comic. The goal was also to learn the various commands and features of the platform.


The following sub-sections describe the process for creating a comic, divided according to the four areas for creating comics within the platform itself: Plan Lesson, Write Script, Layout Storyboard, and Create Final Comic.

\subsubsection{Plan Lesson}
The comic creator has to configure the project for learning objectives; either the creator's own, or those from the NICE Cybersecurity Workforce Framework~\cite{Newhouse2017}.
The comic creator also chooses whether the scores will be saved to the Comic-BEE servers (wherein choices are captured, along with all the relevant data for scoring the comics); or whether readers will be shown their results, in which case the scores are not saved to a server. Optionally, if the creator chooses that scores will be captured, the creator can choose whether to prompt readers with a phrase of his choice; for example, ``Please tell us why this was your choice~:)".
The comic creator also has to define the topic, select what sort of audience will be expected, what kind of learning environment the readers will be in, the knowledge readers should take away, prior knowledge, and the readers' expertise level (before and after exploring the comic). There is also a space here where any known limitations of readers can be enumerated.

\begin{figure}[h]
\centering

\centering
  \includegraphics[width=\linewidth]{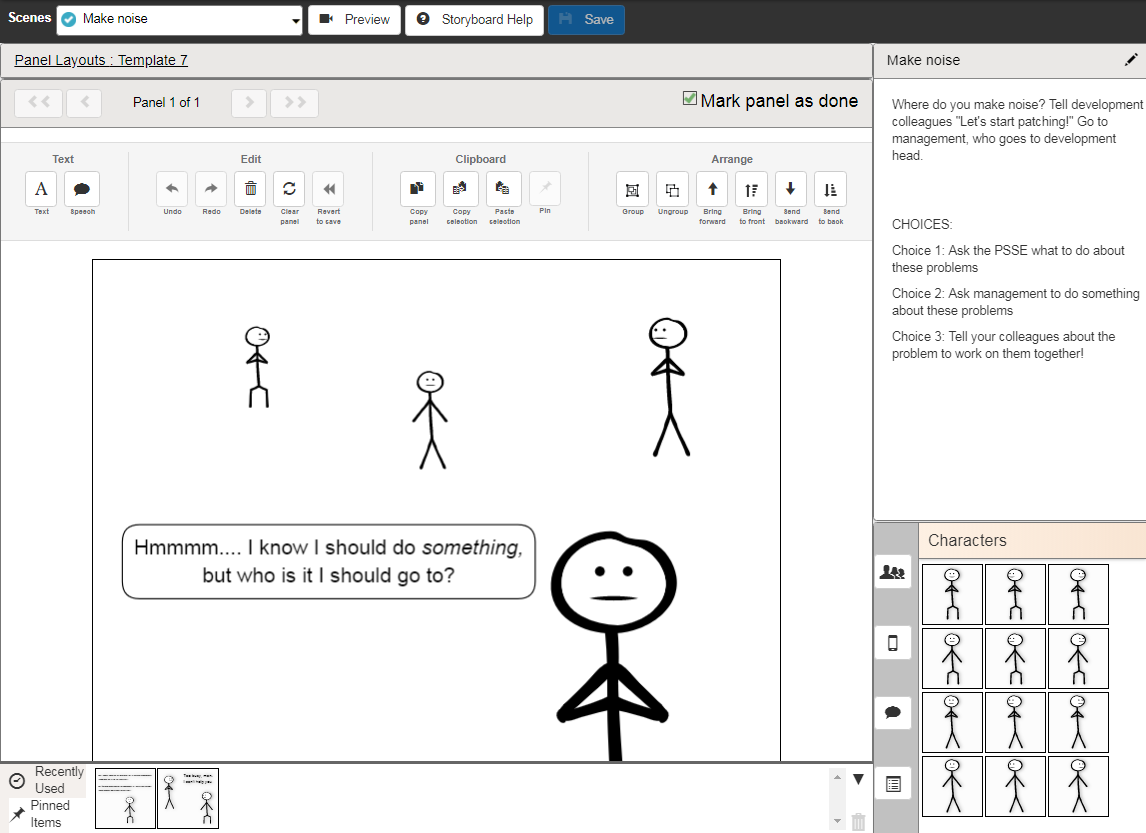}
\captionof{figure}{Black and White working space in \textbf{Layout Storyboard}}
\label{fig:pmstoryboard}
\end{figure}

\begin{figure}[h]
\centering
  \includegraphics[width=\linewidth]{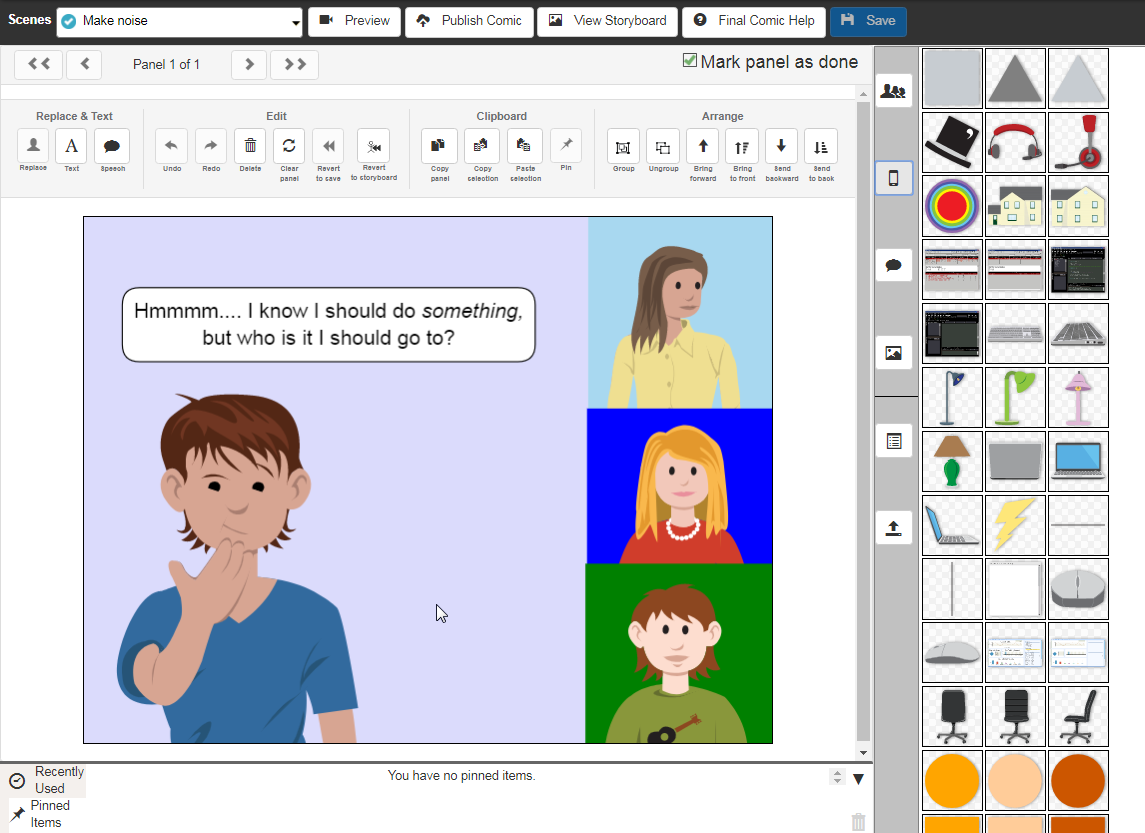}
\captionof{figure}{Full-color working space in \textbf{Create Final Comic}}
\label{fig:pmcfc}

\end{figure}

\subsubsection{Write Script}
\label{subsec:writescript}
In this section, scenes and choices are inputted and organized.
There are options to create scenes based on the Learning Objectives or Real life Scenarios. If the creator selects either of them, a new scene is automatically created that the creator can Edit, Duplicate, or Delete.
Here, the scene is named; designated as Start, Normal, or Ending (different from the Real Life Scenario prompt); the script is written, the question posed; and choices enumerated and rated.
Choices can be rated as five levels of expertise, from Apprentice to Journeyman, and rated from 1 to 5 for quality, Worst to Best.
It is here, in the choices, where specific learning objectives are assigned.
A minimum of one choice is required for Start and Normal scenes, and Ending scenes do not have choices; the maximum number of choices can exceed 20; though, given item choice number theory \cite{Vyas2008,Rodriguez2005,Nwadinigwe2013}, three was found to be sufficient.

\subsubsection{Layout Storyboard}
\label{subsec:cpstoryboard}
The storyboard section is the first for which the creator can edit characters and assets in panels and storyboards. Scenes can be anywhere from 1 to 6 panels, with 8 various arrangements to choose from.
After having selected a panel layout, the creator is given a working space to add any text, items, or characters to each panel individually; to aid in this, the scripts from the Write script section are shown alongside the working space. Characters can be sitting or standing, and be facing  left, forward, or right.
Everything is in black and white, see Figure~\ref{fig:pmstoryboard}.
After having populated all of the panels and scenes, the creator can move on to create the final comic.


\subsubsection{Create Final Comic}
\label{subsec:CFC}
Here, the creator has a working space, just as before, but this time there is a button for replacing the characters and items with their full-color counterparts, see~\autoref{fig:pmcfc}.
Characters can further be customized to have up to ten various expressions,  seven arm positions, four directions to face, and four leg positions.
Items and characters can be stretched, rotated, and mirrored, and they are arranged according to layers within the working space.
To add a sort of variance, background colors can be selected, there are all sorts of shapes, and creators can even upload their own SVG files to be included in their projects.

\subsection{Collecting Learning Material from Experts}
Experts on  information security, organizational procedures, and resource allocation were consulted for this work. 
To begin developing comics, first it was necessary to find topics deemed important by the experts.
We used a questionnaire of three items during phone interviews to find out common problems: 
\begin{enumerate}
 \item[Q1] {What are among the most common problems encountered?}
  \item[Q2] {What are the situations that surround the most common problems? What other choices can be made (either right or wrong)?}
  \item[Q3] {What is the context of  events that occur before or after common problems (do not have to be directly related to the problems themselves), e.g., workload, time-frame, other tasks\ldots{}
}
\end{enumerate}

The questionnaire was designed to allow the respondent to elucidate on all aspects of insecure scenarios, including the proximal and ultimate causes of the problems, the situations that may surround problems, and the contexts surrounding problems that may (seemingly) have little or nothing to do with the problems themselves.

\subsection{Developing Learning Material}
\label{sec:developingscripts}

The final list was comprised of five different learning topics.
In order to create a comic with more branching paths, a second expert (Expert 2) was consulted for creating new ideas that elaborated on the aforementioned topics.
Several methods were tried for collecting and organizing ideas and paths, but the best method seemed to be a visual representation that displayed all the choices, such as the whiteboard, which was used to create the second comic.

\subsection{The Comic Creation Process}
After some trial and error, the following six phases ended up being the most effective method for creating complete and coherent comics.

\begin{enumerate}
	\item \textbf{Design}: Collect ideas from experts and colleagues, and turn them into viable scenes and choices that could be used in a storyboard.
    \item \textbf{Storyboard}: Scenes and choices are organized into a coherent story, especially so they could be included into the comic platform.
    \item \textbf{Comic Platform - Write Script}: The design and storyboard is now taken from the ``page" and put online into the comic platform. 
    \item \textbf{Comic Platform - Storyboard}: The story was arranged into a storyboard. For a full description see Sub-section~\ref{subsec:cpstoryboard}.
    \item \textbf{Comic Platform - Final Draft}: The final draft of the comic was created. For a full description, see Sub-section~\ref{subsec:CFC}.
    \item \textbf{Feedback on Final Draft}: After the comic was fully realized, it would be sent out for feedback once more, and that feedback would be taken to revise the comic one last time before it would be ready to be used. 
\end{enumerate}

\subsection{Developing the Comics}
\label{sec:comicdev}
The five topic ideas were all made into their own dedicated comic, each covering a single topic: Backup and Restore (B\&R), Principle of Least Privilege (PLP), Password Management (PassM), Where to Share Data (WSD), and Patch Management (PatchM).
Altogether, about 60 hours were spent working directly on the comics; indirectly, the  process took approximately 8 weeks. For a breakdown of the amount of time spent working directly on each comic by phase, see \autoref{tab:comictime}.

\input{tables/time.tex}

After completing each phase of the comic creation process, the comics were sent to experts for feedback, whereby the responses were received up to a week later. Each round of feedback consisted of the following five feedback questions for readers:
\begin{enumerate}[]
	\item How realistic are the scenes and choices?
    \item Is anything missing?
    \item Is anything wrong?
    \item Do you think they could work as teaching material? (e.g., in training)
    \item What do you think, in general?
\end{enumerate}
\label{fiveqs}

%% file: tables/time.tex
\begin{table*}
\centering
\begin{threeparttable}
\centering
\caption{Time taken for comic creation (in hours) by phase and comic}
\label{tab:comictime}
\begin{tabular}{
  l
  c c c c c c 
}
\toprule
Phase & B\&R & PLP & PassM & WSD & PatchM & \textbf{Total} \\
\hline
1. Design & 0.75 & 0.75 & 0.75 & 0.5 & 0.75 & \textbf{3.5} \\
2. Storyboard & 2 & 1 & 1 & 1 & 1 & \textbf{6} \\
3. Comic Platform - Write Script & 2 & 1.25 & 0.75 & 0.75 & 1 & \textbf{5.75} \\
4. Comic Platform - Storyboard & 1.75 & 2 & 2 & 2 & 2 & \textbf{9.75} \\
5. Comic Platform - Final Draft & 8 & 4 & 6 & 6 & 6 & \textbf{30} \\
6. Feedback on Final Draft & 1 & 1 & 1 & 1 & 1 & \textbf{5} \\
\textbf{Total} & \textbf{15.5} & \textbf{10} & \textbf{11.5} & \textbf{11.25} & \textbf{11.75} & \textbf{60} \\

\bottomrule
\end{tabular}
\begin{tablenotes}
    \small
     \item \emph{Note:} Feedback was collected for up to 1 week after each phase
    \end{tablenotes}
\end{threeparttable}
\end{table*}

%% file: 4-results.tex
\section{Evaluation}
This section describes the context of the experiment, the participants, and also our results.

\subsection{Setting and Sampling}
Upon completing the comics, the next step was to distribute them among employees of the company, who would evaluate them.
Though this group of employees would be informed that any information they provided would be voluntary and anonymous, the author also preferred that their attention was rapt and that they had an incentive to complete them.
This is why they were not sent out in emails, as was done for the feedback sessions, where it was likely that they would be ignored.
Instead, it was considered better if the comics could be implemented alongside another training, giving the authors a so-called ``captive audience".
Such an occasion was identified in one of the company's many Capture the Flag (CTF) events. The comics were added to the CTF as an exercise to earn points by finding the ``correct" story path leading to a flag that earns points. 

The event lasted two days; on the first day, there were 10 teams of two and one individual participating in the event.
On the second day, there were 5  participants. The participants included Web Developers, personnel from Research and Development, Software Testers, and Project Managers. The participants' age ranged from their 20's to their 50's (see \autoref{tab:pilotparticipants}).

Capture the Flag events are, by their nature, competitive and fast-paced.
Given the structure of the CTF event, participants were given the incentive to complete the comics (and all the other events) for points, and to do so as quickly as possible.
Participants were provided with a virtual machine that contained all of the tasks and objectives; this also allowed the moderators of the event to monitor their progress.
Once an objective was successfully completed, participants were given a random string of digits, whereby these ``flags" could be exchanged for points.
As is typical in a CTF event, the team with the most  points wins the event and is given a small prize.
As the comics were not initially part of the official CTF event, they had to be modified to include flags on the scenes that were considered to be ``correct".

\subsection{Results}
Before the CTF, feedback was mostly positive; the following remarks were typical: ``Very realistic", ``Try mentioning company policy", ``Nothing wrong",  ``It could give an overall idea why backup is important", and ``It is good".
This gives an indication that Requirement 3 is at least partially fulfilled. 
There was one instance of a colleague who found the comics to be indecipherable. He mentioned that he did not know which way to read the comics and even that he did not like comics. This point was noted because it was unexpected that someone would not like comics. However, his feedback does not affect any of our target requirements.

Additional comments that recommended changes were also taken into account, and the corresponding necessary changes were made.

During the CTF events, approximately 20 people participated in the CTF and had viewed the comics.
On the first CTF day, the comics were given to the participants at the beginning of the event in the morning, as a sort of ``warm-up" exercise. Participants needed to solve the comics in order to be able to start with the CTF challenges.
This caused some negative feedback from the participants and some obvious stress; thus clearly indicating that Requirement 2 is not fulfilled.

After the CTF event, which lasted 8 hours, a 10-minute feedback session was held.
The same five feedback questions were asked as were asked to offline participants (see the listing in section~\ref{fiveqs}). Additionally, participants were told that any feedback they provided would be voluntary.
Since distress in the feedback by the participants was noticeable (due to not being anonymous), the feedback strategy was changed for the second CTF day.

For this second session, which took place the following day, the participants were asked to respond to the questions on a slip of paper.
Additionally, they were asked to answer the following demographic questions: gender, age, whether they have read comics before, which kind, whether they like comics, and the field they work in.


On the first day, the author noted that the participants found the comics ``Okay", but that the images themselves were largely ignored, as they were perceived to be unnecessary. One participant said that he would ``just read the answers and make his decision", three other participants agreed with him, and many other participants seemed to concur.
Another participant said that ``[i]t is normally the case that the visuals actually add to the learning experience".
Another participant added, ``In this case it seemed that the pictures were entirely unnecessary, a waste. We were able to skip right over them to read the responses and make our selections without looking at them. What if the answers were also pictures? Because it took me out of the context to be looking at pictures then make a text selection".
By receiving this feedback, requirement 3 and 4 are  obviously not fulfilled.

When asked to volunteer information about their past experience with comics, about 3 of the remaining 10 participants indicated that they had read comics before, and when the author asked who was familiar with the Asterix series all of the participants seemed to be familiar.

The participant feedback for the second CTF day is shown in \autoref{tab:pilotparticipants}. Here we get a glimpse of failed requirements 1 and 4.

On the first CTF event, the participants clearly gave feedback that the comics were perceived as disturbing the CTF game, as they consumed time to answer but did not add to the CTF fun.

\input{tables/pilotparticipants.tex}

%% file: tables/pilotparticipants.tex
\begin{table*}[ht]
\centering
\begin{threeparttable}
\caption{Participant Demographic Information: Day 2}
\label{tab:pilotparticipants}
\begin{tabular}
{
 c c c c c c c 
}
\toprule
\# & Gender & Age Range & Read comics before? & Do you like comics? & Which comics? & What's your field \\
\hline
1 & F & 20's & N & - &- & Software Testing \\
2 & F & 20's & N & N &- & Security Testing \\
3 & M & 40's & Y & Y &Mad & R\&D \\
4 & M & 40's & Y & Y &Asterix & R\&D \\
5 & M & 50's & N & N & - & Web Development \\

\bottomrule
\end{tabular}
\end{threeparttable}
\end{table*}

%% file: 5-conclusion.tex
\section{Conclusions}

This study was conducted to (1) evaluate an online comic-creation platform and (2) to analyze user perceptions of the resultant interactive graphical stories. These stories were implemented for security awareness as part of a capture the flag (CTF) challenge. While the comic creation process was tedious, it was mostly prolonged by the feedback process. The authors thought this could be streamlined, and in a conversation with the platform provider, they have decided to implement a feature that would allow for feedback to be given on the platform itself. Since publication, this new feature has been implemented in an update to the platform.

As for the CTF and feedback, we gained numerous valuable insights.
First, the participants perceived the content of the comics to be realistic. 
Second, stories or vignettes in the form of comics were also considered a good idea for communicating complex ideas, and being used as training material.
The participants perceived the following elements as needing improvement: the images and stories were not seen as essential in the stressful situation (while under time pressure of the CTF) and were therefore, largely ignored.
In particular, during a hacking competition, people tried to apply the hacking mindset to the comics and just tried all paths and possible answers without reading or understanding them. Their only goal was to have the right answer before the other teams, understanding the story was not the main goal. 

Thus, one of our main results is that the comics were not well-received, not because of the comics or the content itself, but rather because of the context in which they were implemented - in a CTF platform.

Based on these insights we have identified the following recommendations:
\begin{itemize}
    \item to save time, feedback can be limited to the initial and final stages of the content creation process
    \item comics, interactive or otherwise, should not be implemented during time-dependent (sensitive) activities such as during a CTF event
    \item it must be ensured that comics are used for training in non-stressful environments, such as during a break or other leisure time, to encourage a relaxed and even playful mood
    \item while they can be about serious matters, the comics can also be fun; we hypothesise that by being fun, they remain in the memory of the participants for longer time
\end{itemize}


Scientific conferences present successful research to a large extent. 
Unfortunately, the investigative journey presented in this work has resulted in our requirements and goal not being achieved.
While these were not the results we were anticipating, it is all the more reason that we are glad to have this work published, so that future researchers can learn from our experiences. In particular, we consider the lessons-learned to be important so that they can be considered when using comics as a form of IT security awareness training. 
